## Theory of laser ion acceleration from a foil target of nanometers

X.Q. Yan[1,2*], T.Tajima[3,4], M. Hegelich[4,5], L.Yin[5], D.Habs[4,1]

[1]Max-Planck-Institut f. Quantenoptik, D-85748 Garching, Germany

[2]State Key Lab of Nuclear physics and technology, Peking University, 100871, Beijing, China, and Center for Applied Physics and Technology, Peking University.

[3]Photo-Medical Research Center, JAEA, Kyoto, 619-0215, Japan

[4]Fakultaet f. Physik, LMU Muenchen, D-85748 Garching, Germany

[5]Los Alamos National Laboratory, Los Alamos, New Mexico 87545, USA

## Abstract:

A theory for laser ion acceleration is presented to evaluate the maximum ion energy in the interaction of ultrahigh contrast (UHC) intense laser with a nanometer-scale foil. In this regime the energy of ions may be directly related to the laser intensity and subsequent electron dynamics. This leads to a simple analytical expression for the ion energy gain under the laser irradiation of thin targets. Significantly, higher energies for thin targets than for thicker targets are predicted. Theory is concretized to the details of recent experiments which may find its way to compare with these results.

## I. Introduction:

The dream of collective acceleration started with the vision of Veksler [1] and Budker [2]. If ions were to be trapped by speeding electron cloud or beam with energy $\varepsilon_e$, the ions would be accelerated to the energy of $\varepsilon_i = (M/m)\varepsilon_e$, where $M$ and $m$ are masses of ions and electrons, respectively, because they would speed with the same velocity. Since the mass ratio $M/m$ of ions to electrons is nearly 2000 for protons and greater for other ions, the collective acceleration of ions would gain a large energy boost. A large body of investigations ensued [3-4]. Also started were investigations of electron clouds to cool and/or accelerate ions in storage ring or traps as a variation of this vision (see, e.g. [5]). In this the electron cloud slightly ahead of the ion beam with some velocity differential can cause a frictional force of the Bethe-Bloch type to drag ions for acceleration and/or cooling, if and when the velocity difference between electrons and ions is controlled under the given condition (an 'adiabatic' condition). The friction force arising from the electron bunch here plays a role similar to the friction played by photon pressure on atoms in the case of S. Chu [6]. None of the collective acceleration experiments, however, found energy enhancement of this magnitude mentioned above. The primary reason for this was attributed to the sluggishness (inertia) of ions and the electrons being pulled back to ions, instead of the other way around, 'reflexing of electrons' as described in [7]. In another word, the fast dynamics of light electrons is mismatched with the slow dynamics of heavy ions. As we shall see in more detail, Mako and Tajima theoretically found that the ion energy may be enhanced only by a factor of $2\alpha+1$ (which is about 6 or 7 for typical experimental situations and $\alpha$ will be defined later in Sec.II) over the electron energy, instead of by a factor of nearly 2000, due to the electron reflexing. (For example, Tajima and Mako[8] suggested to reduce the culpable electron reflexing by providing a concave geometry.) In year 2000 the first experiments [9-12] to collectively accelerate ions by laser irradiation were reported. Since then, a large amount of efforts have been steadily dedicated to this subject.

[*] Email: xyan@mpq.mpg.de

Because of the advantage in accelerating limited mass by laser to cope with the mismatch between the electron and ion dynamics as discussed above, experiments producing high-energy ions from sub-micrometer to nanometer targets much thinner than ones in early experiments driven by ultrahigh contrast (UHC) short-pulse lasers [13–17] have attracted a recent strong interest. Of particular focus is how much the ion energy enhancement is observed in the experiments and simulations in these thin targets and how it scales with the laser intensity.

The experiments and simulations of late show that the proton energy increases as the target thickness decreases for a given laser intensity, and that there is an optimal thickness of the target (at several nm) at which the maximum proton energy peaks and below which the proton energy now decreases. This optimal thickness for the peak proton energy is consistent with the thickness dictated by the relation $a_0 \sim \sigma = \dfrac{n_0}{n_c}\dfrac{d}{\lambda}$, where $\sigma$ is the (dimensionless) normalized electron areal density, $a_0$, $d$ are the (dimensionless) normalized laser amplitude and target thickness [18-20]. This is understood as arising from the condition that the radiation force pushes out electrons from the foil layer if $\sigma \leq a_0$ or $\xi \leq 1$, while with $\sigma \geq a_0$ or $\xi \geq 1$ the laser pulse does not have a sufficient power to cause maximal polarization to all electrons. Here we have introduced the dimensionless parameter of the ratio of the normalized areal density to the normalized laser amplitude $\xi = \sigma / a_0$. Note that this optimal thickness for typically available laser intensity is way smaller than for cases with previously attempted target thicknesses (for ion acceleration). Thus we attribute the observed singularly large value of the maximum proton energy in the recent experiment [21] to the ability to identify and provide prepared thin targets on the order of nm to reach this optimal condition. In reality at this target thickness the laser field teeters over partial penetration through the target, rendering the realization of optimum rather sensitive. Under this condition, electron motions maintain primarily those organized characteristics directly influenced by the laser field, rather than chaotic and thermal motions of electrons resulting from laser heating. In 1D PIC simulation (Fig.1) we observe that momenta of electrons show in fact coherent patterns directing either to the ponderomotive potential direction, the backward electrostatic pull direction, or the wave trapping motion direction, in a stark contrast to broad momenta of thermal electrons. In another word, through a very thin target the partially penetrated laser fields enable the electrons to execute dynamic motions still directly tied with the laser rather than thermal motions. Furthermore, since expelled electrons form a dense plasma sheet whose density can exceed $n_c$, some fraction of the penetrated laser pulse seems to be trapped between the diamond foil and the newly formed electron sheet, as we discuss in Sec.IV. We note that the ponderomotive force due to this trapped radiation contributes to the acceleration of electrons in this sheet and thus retards these electrons from being decelerated by the electrostatic force emanated from the diamond foil. In a typical sheath acceleration scheme the termination of ion acceleration commences due to this electron reflexing by the electrostatic field.

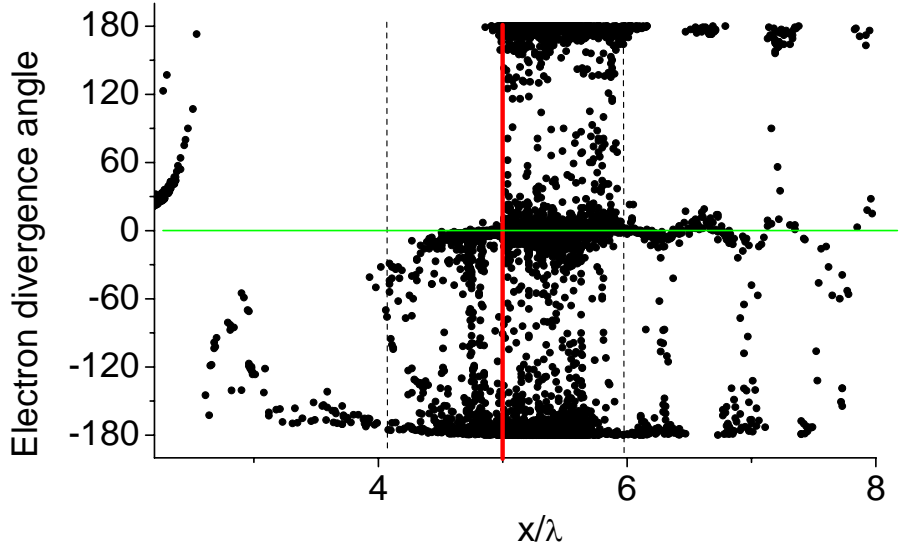

Fig.1 Coherent electron motions in the laser irradiation on a thin target. The electron divergence angle $\tan^{-1}(p_y/p_x)$ versus position $x$ at $t = 22$, where space $x$ is measured in the wavelength $\lambda$ and it is in the direction of laser propagation and $y$ is the polarization direction, while time $t$ is normalized by the laser cycle. On the left of the target we see electrons backwardly spewed out near angle of -180°. In the forward direction we see forward electrons at 0° due to the ponderomotive force, and electrons reflexing by the electrostatic fields (180° or -180°). We further see electrons trapped in some wavelike structure, changing swiftly their directions. All these are indicative of the direct imprint of the electron motion in the laser fields. Note also that even within the target we discern structured electron loci, showing electrons driven by some minute structured (perhaps the wavelength of $2\pi c/\omega_p$) fields in the target. (Laser amplitude a=3.6, $\xi \sim 1$ and normal incidence. The vertical bold line represents the initial target located at $x = 5\lambda$, two dotted lines show the boundaries of the expanded target.)

On the other hand, most of the theories have been based on the so-called Plasma Expansion Model (PEM) [22], which is motivated by much thicker and massive target. In this regime electrons are first accelerated by the impinging relativistic laser pulse and penetrate the target driven by ponderomotive force. Leaving the target at the rear side, electrons set up an electrostatic field that is pointed normal to the target rear surface, which is the so-called TNSA (Target Normal Sheath Acceleration) acceleration. Most electrons are forced to turn around and build up a quasistationary electron layer. These fast electrons are assumed to follow thermal or Boltzmann distribution in theoretical studies of the conventional TNSA mechanism [13,22-24], where the acceleration field is estimated by the exponential potential dependency in the Poisson equation. Though this mechanism is widely used in the interpretation of the experimental results, it does not apply to the ultrathin nanometer scale targets, because the direct laser field and attenuated partially transmitted laser pulse play an important role in electron dynamics and the energetic electrons oscillate coherently, instead of chaotic thermal motions. Based on a self-consistent solution of the Poisson equation and TNSA model, Andreev et al [13] had proposed an analytical model for thin foils and predicted the optimum target thickness is about 100 nm. It obviously does not explain the experimental results [21].

In Secs.II and III we formulate the dynamic treatment of electrons and ions, respectively, coupled through the electrostatic potential self-consistently for laser irradiated ion acceleration from a thin target. In Sec.IV we discuss the physical effects other than the direct imprints of laser fields on particles after laser goes through the target. In Sec.V we consider physical processes that become relevant when the target is thicker ($\xi \gg 1$), but not on a scale of thickness in the conventional TNSA. In Sec.VI we make our conclusions.

## II. Electrostatic potential in coherent dynamics

We formulate the maximal ion energies in the laser driven foil interaction of our regime in this paper, without assuming thermalized electrons. When the foil is thick with $\xi \gg 1$ and the laser pulse is completely reflected, the ion acceleration may be described by the plasma expansion model for TNSA [22]. In the contrary, in case of $\xi \ll 1$, the transmission is dominant and the laser passes without too much interaction with the target. However, we will note that there is a regime ($\xi \gg 1$) with thickness still much smaller than that for TNSA (to be discussed in Sec.V).

The optimum ion acceleration condition is, as discussed, in the range of $\xi \sim 1$ ($0.1 < \xi < 10$). There appears partially transmitted laser pulse and behind the target energetic electrons still execute the collective motions in the laser field. Electrons quiver with the laser field and are also be pushed forward by the ponderomotive force. We see in Fig.1 that in the region ahead of the exploding thin target, there are three components of characteristics orbits: a set of orbits in forward direction with angle 0°), the second backward (with -180° or 180°), and the third with loci with curved loops. The first two are characteristics observed even in a simple sheath, but also present in the current case, where perhaps the forward is as vigorous or more so as the backward one. The third category belongs to the orbits of trapped particles in the laser field or the ponderomotive potential. For a reflexing electron cloud the distribution shows only two components, the forward one and the backward one.

In an ultra-thin target, the laser electromagnetic fields largely sustain coherent motions of electrons. As partially penetrated laser fields in addition to the laser fields in the target, the electron motion under laser fields is intact and is characterized by the transverse field. The electron energy is consisted of two contributions, the kinetic energy of (organized) electrons under laser and the ponderomotive potential of the partially penetrated laser fields that help sustain the electron forward momentum. We discuss these aspects in more detail in Sec.IV. Following the analysis of Mako and Tajima [7], the plasma density can be determined by:

$$n_e = 2 \int_0^{V_{max}} g(V_x) dV_x ,$$ (1)

$$V_{max} = c \sqrt{1 - m_e^2 c^4 / (E_0 + m_e c^2)^2} ,$$ (2)

where $g$ is the electron distribution function and $E_0$ is the characteristic electron energy.

The forward current density of electrons $J$ and electron density $n_e$ are related through

$$J = -e \int_\upsilon^{V_{max}} V_x g dV_x ,$$ (3)

$$n_e = \frac{2}{e} \int_0^{V_{max}} \frac{dJ/d\upsilon}{\upsilon} d\upsilon .$$ (4)

At a given position in the reflexing electrons cloud where the potential is $\phi$, the total particle energy is given by

$$E = (\gamma - 1)m_e c^2 - e\phi .$$ (5)

Current density can be determined from the simulations results. We find that the current density dependence on $E$ is not exponential, but rather well fits with a power-law. (The origin of such relationship may arise from electrons in our regime retaining coherent dynamics, rather than put into an equilibriating thermal motions). The power-law dependence may be characterized by two parameters, the characteristic electron energy $E_0$ and the exponent of the power-law dependence on energy $E$.

$$J(E) = -J_0(1 - E/E_0)^\alpha. \tag{6}$$

The index $\alpha$ here designates the steepness of the energy dependence on electrons and is a measure of coherence of the electron motion. In another word the greater $\alpha$ is, the more electrons in coherence motion are contributing to the overall current of electrons. Thus we may call $\alpha$ as the coherence parameter of electrons. Usually the most energetic electrons are lost from the system and have minor contribution to the ion acceleration [25-27]. The maximum electrostatic potential is smaller than the laser ponderomotive potential or the characteristic electron energy $E_0$. In the high laser intensity case the relativistic electrons are dominant so that the integral is carried out with the relativistic kinematics as:

$$n_e = \frac{2}{e} \int_0^{V_{max}} \frac{dJ/dv}{v} dv = \frac{2}{ec} \int_{-e\phi}^{E_0} \frac{dJ}{dE} dE = -\frac{2J_0}{ec}(1 + e\phi/E_0)^\alpha = n_0(1 + e\phi/E_0)^\alpha. \tag{7}$$

Fast electrons do not contribute to the acceleration of ions (both proton and carbons). This is why these electrons with lower energies are used to measure the exponent $\alpha$ for *(-J)* versus *E*. Fig.2 shows two snapshots of J(E) curves from PIC simulations, where theoretical curves are plotted at t=18 and at t=24). Since the maximum laser energy transmission to electrons and the potential energy (the sum of these two) at the time when the laser pulse has just transmitted the target, we measure the functional relationship between *-J* and *E*. In Fig.2(a), we superpose this function with the choice of $E_0 = 6.3$, $\alpha = 3$ in the expression of Eq.(6). In another word, the maximum characteristic electron energy is reaching 6.3 in the unit of $mc^2$ at the end of laser plasma interaction, meanwhile the exponent $\alpha$ remains approximately constant around 3 during this phase of evolution under the optimum ion acceleration condition: $\xi \sim 1$. Fig.2 shows $\alpha$ is not very sensitive with $\xi$ varying from 0.2 to 5, while it drops down from this range.

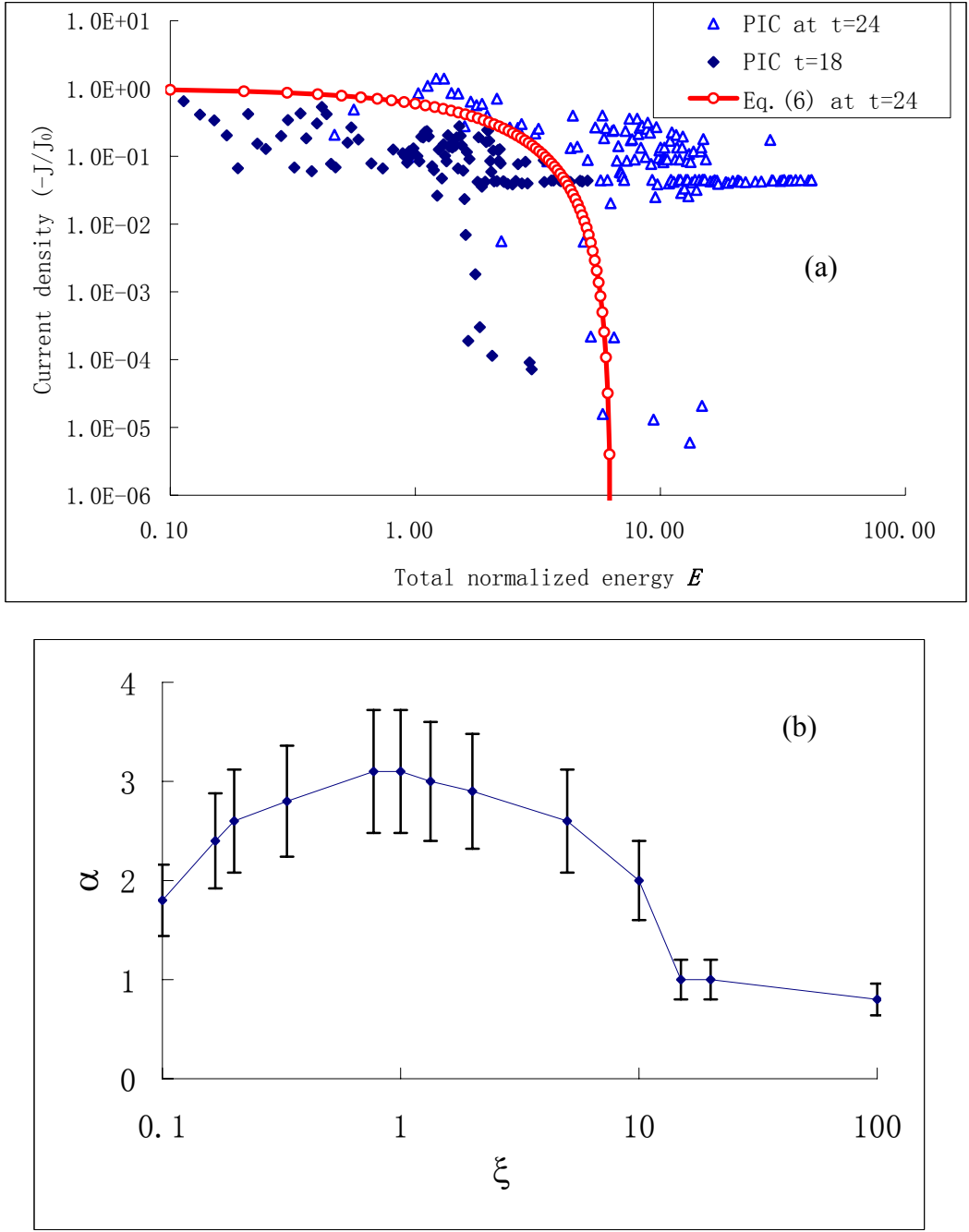

Fig.2 (a)Backward electron current density (-J) versus electron total energy $E$ from PIC simulation with $a_0 = 7.2$ (Typically the highest energy transfer of laser to the electron and potential energies at the time of laser pulse having transited. In this example, it is around t=24, when we find the values of $E_0 = 6.3$, $\alpha = 3$, which are defined in Eq.(6). It is noted that because of statistical fluctuations and some ejected high energy electrons, we encounter relatively large fluctuations away from this curve); (b) a plot of the coherence parameter $\alpha$ versus the normalized areal density $\xi$ (keeping $\xi \sim 1$, the simulation box is the same as Fig.1).

# III. Self-similar evolution of ion dynamics

The system's evolution needs to be tracked self-consistently with electrons, ions and the interacting electrostatic potential in time. These consist of highly nonlinear coupled system of equations. We treat electrons as discussed in section II, while we describe ions in non-relativistic nonlinear equations in this section.

The non-relativistic fluid equations are used to describe the response of the ions to the electrostatic field as follows.

$$\frac{\partial n_i}{\partial t} + \frac{\partial}{\partial x}(\upsilon_i n_i) = 0 \,, \tag{8}$$

$$\frac{\partial \upsilon_i}{\partial t} + \upsilon_i \frac{\partial \upsilon_i}{\partial x} = -\frac{Qe}{M}\frac{\partial \phi}{\partial x} \,. \tag{9}$$

In order to solve the equations self-consistently, the self-similar condition is invoked by using the fluid equations and electron distribution with the self-similar parameter

$$\zeta = x/(\upsilon_0 t) \,, \tag{10}$$

$$\upsilon_0 = (Qe\phi_0/M)^{1/2} \,, \tag{11}$$

$$e\phi_0 = E_0 \,, \tag{12}$$

which is the characteristic electron energy. We introduce the dimensionless parameters:

$$U = \upsilon_i/\upsilon_0, \, \Re = n_i/n_0, \, \psi = \phi/\phi_0 \tag{13}$$

Eq. (8) and Eq.(9) now take the form of:

$$\Re'(U-\zeta) + \Re U' = 0 \,, \tag{14}$$

$$U'(U-\zeta) + \frac{d\psi}{d\Re}\Re' = 0 \,, \tag{15}$$

$$\Re = (1+\psi)^\alpha \tag{16}$$

The conservation of energy is assessed with the boundary condition on the surface of the target:

$$U^2/2 + \psi = 0 \quad \text{at} \ \ \zeta = 0 \,. \tag{17}$$

The solutions to the set of Eqs.(14-16) are:

$$\Re = \left\{ \frac{\alpha}{(2\alpha+1)^2} \left( \zeta - \sqrt{2(2\alpha+1)} \right)^2 \right\}^{\alpha}, \tag{18}$$

$$U = \frac{2\alpha+2}{2\alpha+1} \zeta - \sqrt{\frac{2}{2\alpha+1}}, \tag{19}$$

$$\psi = \frac{\alpha}{(2\alpha+1)^2} \left( \zeta - \sqrt{2\alpha+1} \right)^2 - 1. \tag{20}$$

Eqs. (18-20) also read in usual units as:

$$n_i = n_0 \left\{ \frac{\alpha}{(2\alpha+1)^2} \left( \zeta - \sqrt{2(2\alpha+1)} \right)^2 \right\}^{\alpha}, \tag{21}$$

$$\upsilon_i = (\frac{QE_0}{M})^{1/2} \left( \frac{2\alpha+2}{2\alpha+1} \zeta - \sqrt{\frac{2}{2\alpha+1}} \right), \tag{22}$$

$$\phi = \phi_0 \frac{\alpha}{(2\alpha+1)^2} \left( \zeta - \sqrt{2\alpha+1} \right)^2 - \phi_0. \tag{23}$$

The maximum energy is assessed when the ion density vanishes. This yields from Eq.(18-19):

$$\varepsilon_{\max,i} = (2\alpha+1)QE_0 \quad . \tag{24}$$

In Eq.(24) we see that the ion energy is greater if the coherence parameter of electrons is greater. This tendency is thought of as a legacy of Veksler's vision, i.e. if and when electrons would "behave OK" (i.e. coherent), the energy of ions could be greater.

A more general expression for the time-dependent maximum kinetic energy at the ion front from Eq.(22) is:

$$\varepsilon_{\max,i}(t) = (2\alpha+1)QE_0((1+\omega t)^{1/2\alpha+1} - 1), (t \leq \tau). \tag{25}$$

Here $\tau$ is the laser pulse duration and $\omega$ is the laser frequency. At the beginning the ion energy $\varepsilon_{\max,i}(0) = 0$ and the ion energy approaches infinity as long as the time $t \to \infty$. Normally as the maximum pulse duration of a CPA (Chirped Pulse Amplification) laser is less than pico-seconds, the final ion energy from Eq.(25) is only about $\varepsilon_{\max,i}(t=1\text{ps}) = 2(2\alpha+1)QE_0$. This means that the long pulse is not so advantageous for the energy enhancement of ions. We will discuss the ion energy dependence on the pulse length in detail in Sec.V.

# IV. Secondary Processes after the Laser Impingement

When a short laser pulse impinges on a very thin target ( $\xi \lesssim 1$ ), we notice some interesting phenomenon. The laser pulse can partially penetrate the target and enables the electrons to execute dynamic motions still directly tied with the laser rather than thermal motions, as we discussed. Fig.2 shows the expelled electrons form a plasma sheet whose density can exceed $n_c$. Some fraction of the penetrated laser pulse seems to be trapped between the diamond foil and this newly formed electron sheet. Therefore, this means that ions are affected not only by the electron energy $E_0$, but also the ponderomotive potential $\Phi_{pt}$, which can push electrons less suddenly than the original laser interaction on the target. Thus we need to introduce the characteristic electron energy as:

$$\varepsilon_{max} = E_0 + \Phi_{pt} .$$ (26)

Because the electron motions are still coherent in the laser field, the electron energy and ponderomotive potential energy can be estimated directly by the laser intensity as[28]:

$$E_0 = m_e c^2 (\gamma_p - 1), \Phi_{pt} = m_e c^2 (\gamma_{pt} - 1) ,$$ (27)

$$\gamma_p = \sqrt{1 + (1-T)a_0^2} ,$$ (28)

$$\gamma_{pt} = \sqrt{1 + T a_0^2} ,$$ (29)

where $\gamma_p$ and $\gamma_{pt}$ are the associated electron kinetic energies for incident pulse and transmitted laser pulse. The transmission coefficient $T$ in case of the normal incidence can be estimated by [29]:

$$T = \frac{1}{1 + (\pi \xi)^2} .$$ (30)

The maximum ion energy is evaluated by Eq.(31) at time $t = \tau$. The theoretical predication of maximum ion energy happens at around 4nm, which corresponds to the optimum condition $\xi \sim 1$ as Fig.4 shows:

$$\varepsilon_{max,i} = (2\alpha + 1)Q[(\sqrt{(1-T)a_0^2 + 1} - 1) + (\sqrt{Ta_0^2 + 1} - 1)] . \tag{31}$$

We note here that the setup of the electron sheet and consequential ponderomotive potential buildup as we discussed here no longer occur if the foil is thicker ( $\xi \gg 1$ ). Therefore, in the thick regime the split of two terms as expressed in Eqs(26), (27), and (31) do not arise and we should simply take Eq.(24). It is also important to notice that in Fig.3 the transmitted laser pulse has changed its property from that of the incident one, showing much more minute structure of the field oscillations. Detailed analysis shows that the transmitted laser pulse consists of the fundamental laser frequency as well as higher harmonics generation (HHG) from order 2,3,4, and so forth to substantial orders. They are both below and above the cutoff frequency of $\omega_p$ in the target. (We will report these features of HHG in a separate future paper). It should be emphasized that this transmitted low order HHGs are singular and specific to the organized coherent electron motion directly driven from the electromagnetic fields of the laser penetrated trough the thin target.

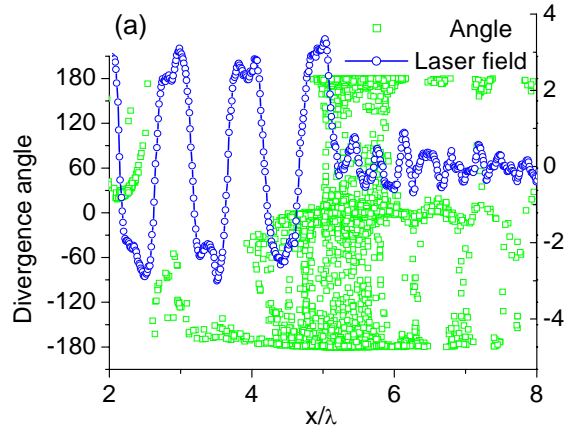

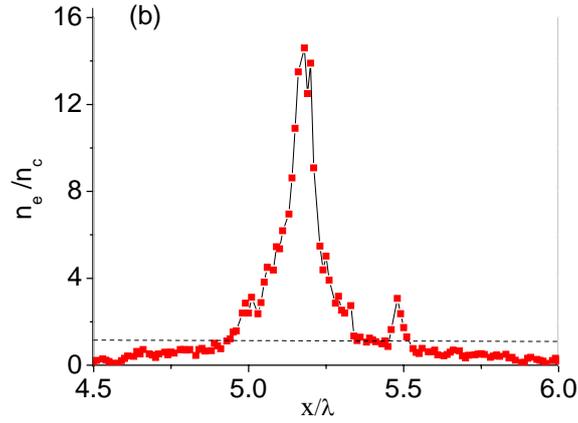

Fig.3 Snapshot of the laser fields and the display of electron divergence angles at t=22 (a) The blue circles are the laser field and the green squares the electron divergence angles. This shows laser is partially transmitted through the target and trapped between the two electron layers; (b)Behind the target there is the second electron layer formed due to the laser ponderomotive push. (Simulation parameters are the same as Fig.1)

We now discuss more detailed processes regarding cases with the interaction of a short laser pulse. To elucidate these processes, we show 2D simulation results. In the simulation runs nanometer scale DLC targets [30] are studied. This may be modeled to have a rectangular shaped plasma with an initial density of $n_e \cong 500 n_c$ consisting of 3 ion species ($C^{6+}$, $C^{5+}$ and $H^+$) in the number ratio of 1/2 : 1/4 : 1. The 2D-PIC code uses a trapezoidal LP shape with a FWHM of 18 laser cycles and a duration of 50 fs, including a rise time of 2.8 fs, a spot size of 6μm and an intensity of $2.6 \times 10^{19}$ W/cm$^2$. The laser wavelength is 0.8 $\mu$m. These theoretical and simulation results are used to interpret the recent experiments [21]. Our theory well predicts the simulation results and outlines these experiments.

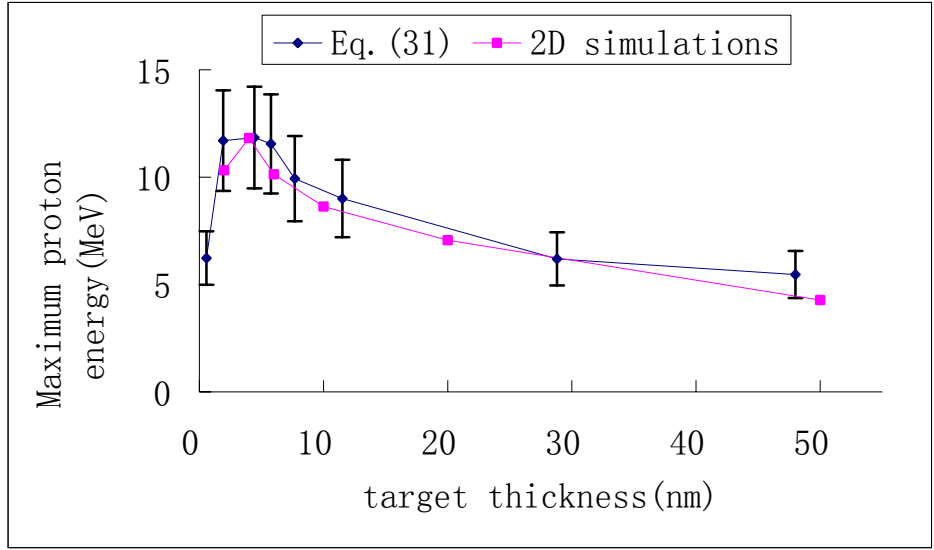

Fig.4. Maximum proton energy versus the target thickness with the laser intensity kept constant at $2.6 \times 10^{19}$ W/cm$^2$ and the pulse length at 50fs. [Notice that $\alpha(\xi)$ is thickness dependence in Eq.(31). See Fig.2(c)].

In the second series of runs, we try to study the maximum proton energy by varying the different laser pulse duration, while keeping the laser energy as constant (at 5J). The LP laser pulse has a Gaussian profile in both the transverse and longitudinal directions with the spot size of $0.8\mu m$ and the same kind of DLC targets are used in the simulations. Eq.(31) is evaluated at $t = \tau$ to compare these simulations. The maximum energy versus the pulse duration is plotted in Fig.5. This shows that a shorter pulse is more favorable to enhance the proton energy. The obtained ion energies are quite impressive. Again we find that the theory well indicates the expected values of energies that we produced from 2D simulations.

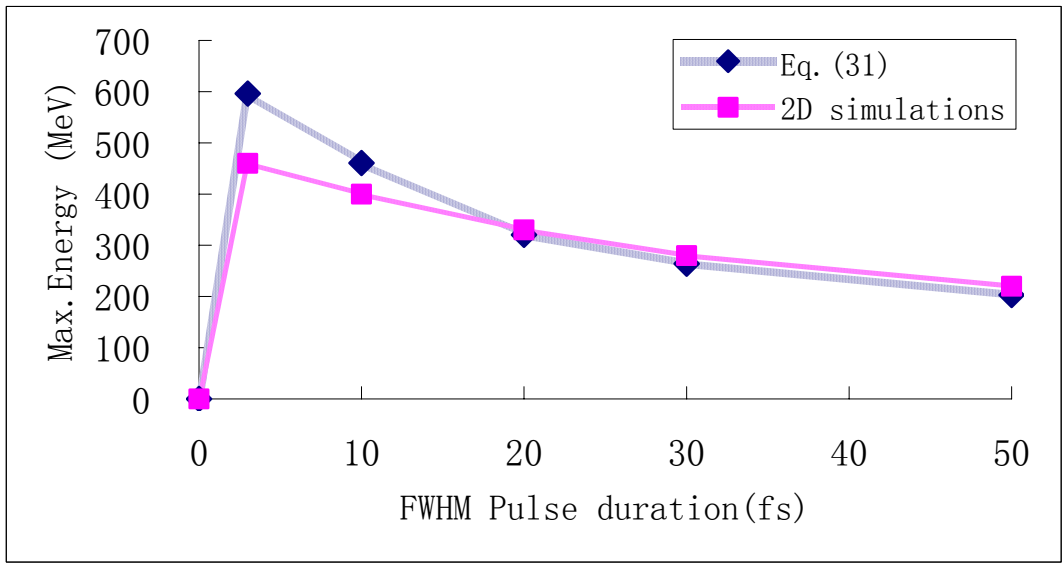

Fig5. Maximum proton energy versus the pulse duration (in the optimal condition $\xi \sim 1$ and $\alpha \cong 3$) with the laser energy kept constant at 5 joules. The peak is reached at 3fs, a single oscillation pulse. The target thickness is taken as 300nm.

# V. Relativistic Transparency and Burn-through

We now proceed to consider the cases when the target is thicker ($\xi \gg 1$) than when it is immediately influenced by the laser fields. In this case the laser does not immediately penetrate through the target. When the target becomes thick so that $\xi$ becomes much greater than unity, our model we discussed in Sec.III becomes less robust, to which we need to remain cautious about its applicability. In this case the interaction process is more complex and we realize that we can delineate at least three stages. The first stage is similar to the situation in Section III. The laser just impinges on the thin surface layer of the dense target. The second stage is after the target begins to expand by the laser interaction primarily in the direction of laser propagation until the laser becomes relativistically transparent at time $t_1$. After this relativistic transparency $t_1$, the plasma expands in all three dimensions. The third stage begins when the plasma becomes underdense at time $t_2$ till the pulse is over. (Here we have assumed a case where the pulse length is greater than both $t_1$ and $t_2$ for the sake of concreteness).

Now we wish to evaluate the plasma expansion in terms of the two characteristic times $t_1$ and $t_2$ as discussed above. In the solid density plasma the skin depth is so short that the ponderomotive force is opposed by the charge separation force beyond. Therefore, the foil expansion in the longitudinal direction may be written as Eq.(32), where a laser pulse with a profile $a = a_0 \sin^2(\Omega t)$ is assumed ($\Omega = \frac{\pi}{2\tau}$), as

$$\frac{dp}{dt} = \frac{Qe\phi}{x} = \frac{Qe\phi_0\sqrt{a_0^2+1}}{x}\sin^2(\Omega t) ,  \qquad (32)$$

$$xdx = \frac{Qe\phi_0\sqrt{a_0^2+1}}{M}(t-\frac{1}{2\Omega}\sin(2\Omega t))dt ,  \qquad (33)$$

$$x^2 - d^2 = \frac{m}{M}\frac{Qc^2\sqrt{a_0^2+1}}{3}\Omega^2 t^4 .  \qquad (34)$$

Assuming expansion only in the x direction at the relativistic transparency the expanded distance $x_1$ may be evaluated by $x_1 = Nd / \gamma = Nd / \sqrt{a_0^2 + 1}, x_1 >> d$.

Then we obtain:

1) 1D expansion time $t_1$

$$t_1 = (\frac{M}{m} \frac{3N^2 d^2}{Q\Omega^2 c^2 a_0^3})^{1/4} = (\frac{12}{\pi^2} \frac{M}{Qmc^2} \frac{N^2 d^2}{a_0^3} \tau^2)^{1/4} \cong (\frac{12}{\pi^2})^{1/4} \frac{N^{1/2}}{a_0^{1/2}} (\tau d / C_s)^{1/2}. \qquad (35)$$

Here $C_s \cong (Qmc^2 a_0 / M)^{1/2}$. The relativistic transparency time $t_1$ in Eq.(35) is in the ball park of the geometrical mean of the laser pulse length $\tau$ and the traverse time over the target by the sound speed. During this period, the laser pulse penetration is limited as expressed by the transmission coefficient Eq.(30). Thus when we integrate the impact on the electron energy at the rear surface of the target to evaluate the $E_0$, we need to incorporate this effect.

$$\bar{E}_0(t_1) = m_e c^2 \int_0^{t_1} (\sqrt{T(t')a^2(t') + 1} - 1) \frac{dt'}{t_1}. \qquad (36)$$

$$\varepsilon_{max,i}(t_1) = (2\alpha + 1)Q\bar{E}_0(t_1)((1 + \omega t_1)^{1/2\alpha + 1} - 1). \qquad (37)$$

This integral $I$ in Eq.(36) may be evaluated if we split this into two pieces, the contributions $I = I_1 + I_2$ arising from $t = 0, t_1 - \Delta t_1$ and that from $t = t_1 - \Delta t_1, t_1$, where $Nc\Delta t_1 /(\lambda a_0) = 1$. The first term may be evaluated as

$$I_1 = m_e c^2 \int_0^{t_1 - \Delta t_1} (\sqrt{T(t')a^2(t') + 1} - 1) \frac{dt'}{t_1} \cong m_e c^2 \frac{\overline{a^2(t_1 - \Delta t_1)}}{\pi(Nd / \lambda)}, \qquad (38)$$

While the second integral may be estimated as

$$I_2 = m_e c^2 \int_{t_1 - \Delta t_1}^{t_1} (\sqrt{T(t')a^2(t') + 1} - 1) \frac{dt'}{t_1} \cong m_e c^2 a(t_1) \frac{\lambda}{ct_1}. \qquad (39)$$

Both of these terms in Eqs.(38) and (39) are multiplied by a coefficient typically much smaller than unity over the expression equivalent to Eq.(24).

Yin et al.[32] have found that for irradiation with a long pulse laser exhibits an epoch of laser burn-through or "breakout afterburner" (BOA). This phenomenon is when the laser goes through a target and eventually it emerges from the rear end of the target. This corresponds precisely to the second period between $t_1$ and $t_2$. We now characterize physical processes including these phenomena. Beyond time $t_1$ the plasma is relativistically transparent so that the laser can now interact with the (expanded) target plasma in its entirety. It can also now expand in three dimensions. In Fig.6 we compare this theoretical value of $t_1$ and $t_2$ with these simulations. For 3D spherical isotropic expansion, it takes time $\Delta t$ during which the normalized density reduced from $\gamma$ to 1:

$$x_2^3 = x_1^3 \gamma \,. \tag{40}$$

That is

$$dt = \frac{dx}{C_s(t)} = \frac{dx}{(qemc^2 a_0 \sin^2(\Omega t)/M)^{1/2}} = \frac{dx}{C_s \sin(\Omega t)} \,. \tag{41}$$

We obtain.

$$\Delta t = \frac{x_2 - x_1}{C_s} \frac{1}{\sin(\Omega t_1)} = \frac{x_1(\gamma^{1/3} - 1)}{C_s} \frac{1}{\sin(\Omega t_1)} = \frac{Nd(\gamma^{1/3} - 1)}{\gamma C_s} \frac{1}{\sin(\Omega t_1)} \,. \tag{42}$$

Now the time $t_2$ when the plasma becomes underdense is given as:

$$t_2 = \Delta t + t_1 \,. \tag{43}$$

In Fig.6 we compare this theoretical value of $t_1$ and $t_2$ with these simulations.

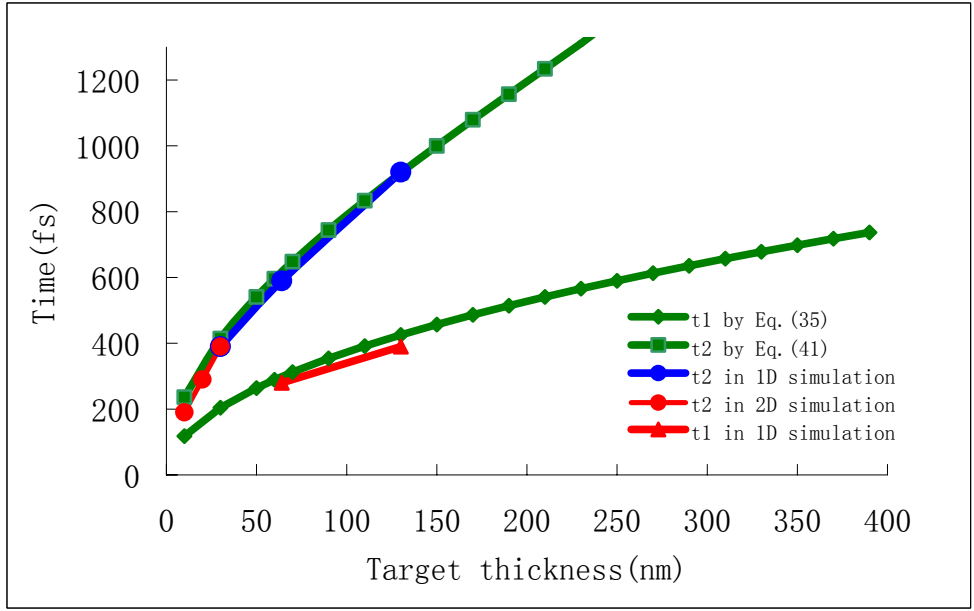

Fig.6 The relativistic transparency time $t_1$ and the burn through time $t_2$ as a function of the target thickness. We compare our theory with 1D and 2D simulations [33]. (In the simulations $a_0 = 20$, $\lambda = 1 \mu m$, the FWHM pulse duration $\tau = 700 fs$, the plasma with an initial density of $n_e \cong 800 n_c$ consists of 2 ion species (C$^{6+}$ and H$^+$)).

Now as we examine the physical situation, we realize that at time $t_1$ the laser pulse has penetrated the entire target with the relativistic transparency and we may regard that the laser begins to drive the entire plasma electrons from this already expanded target. This process may be once again regarded to evolve in a self-similar fashion. If and when this is the case, the slab of plasma that has been penetrated with the laser may be subject to an expansion in a self-similar treatment similar to what we have done for time before $t_1$. (A slight difference remains in that we regard our self-similar expansion is triggered at the laser impingement before $t_1$, while the second self-similar expansion may be considered to commence at the rear end of the target after $t_1$.) With this picture we may exercise the same mathematical tracking of the electrons, ions, and their electrostatic field between them assuming the self-similarity initiated at $t_1$. This process may well be three dimensional, as we discussed. Here, however, for simplicity sake, if we take the same one-dimensional self-similar treatment, we are led to an expression in a closed form for the ion energy gain between time $t_1$ and $t_2$ in the case of a laser pulse with the duration longer than the characteristic time $t_1$ as:

$$\varepsilon_{\max,i,BOA} = (2\alpha+1)Q\overline{E}_0((1+\omega(t_2-t_1))^{1/2\alpha+1}-1) . \tag{44}$$

Here $\overline{E}_0$ is evaluated over time interval $(t_1, t_2)$ and also note that after $t_1$ transmission $T$=1. We have assumed that $t_1 < 2\tau$ and $t_2 < 2\tau$. As we remaked, Eq.(41) has been derived for one dimensional self-similarity. It is thus considered that this would yield an overestimate of energy than a fully three dimensional solution.

Taking these expressions in Eq.(37) and Eq.(44), when $t_1, t_2 < 2\tau$, the total ion energy gain can be obtained. In Fig.7 we plot the total energy gain in the case of carbon from this formula as a function of the target thickness, while the contributions from $\varepsilon_{\max,i}(t_1)$ and $\varepsilon_{\max,i,BOA}$ are also shown. It shows the BOA term is dominant for thick targets. When this condition $t_1, t_2 < 2\tau$ is not fulfilled, an appropriate corresponding modification for energy is due.

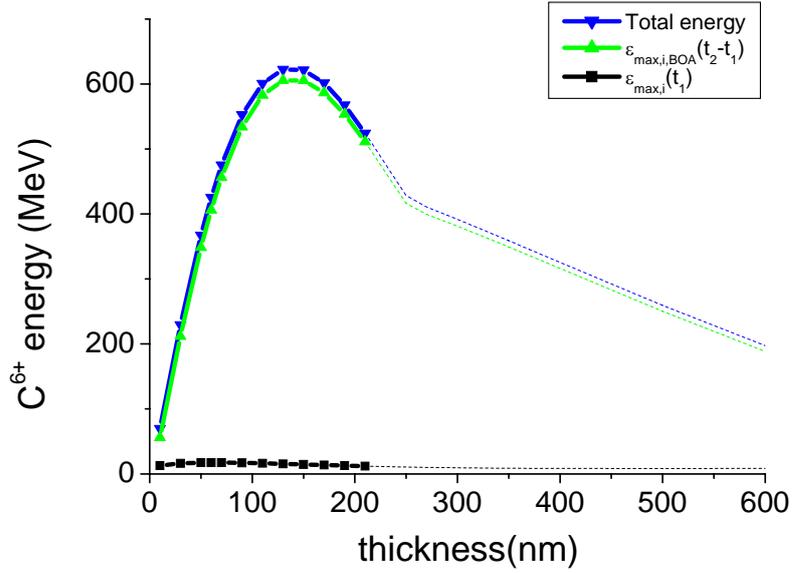

Fig.7 $C^{6+}$ energy gain estimated from Eqs. (37) and (44) as a function of the target thickness and with $\alpha \cong 3$. Beyond $\xi > 10$, where $\alpha$ is supposed to quickly decrease and the model's predictiveness decreases (For a given laser pulse length at 700fs and laser amplitude $a_0 = 20$). The contribution of electron energy gain during the breakout afterburner epoch is dominant.

It should be further noticed that the laser ponderomotive force can also transversely shove electrons and thereby ions over time time $t_t$, while it longitudinally expands the foil plasma at the same time. We evaluate this time $t_t$ in the same way as $t_1$ and it reads:

$$t_t \cong (\frac{24}{\pi^2})^{1/4}(\tau r_0 / C_s)^{1/2}. \tag{45}$$

Here $r_0$ is the laser spot size. In many of our applications it may be that the shoving time $t_t$ is greater than the time scales $t_2$ and $t_1$ for the nm-scale foils. However, sometimes $t_t$ may become smaller than $t_1$, depending on parameters.

# VI. Conclusion:

When a cloud of energetic electrons are injected into a space delineated by a foil of metal or plasma, a strong electrostatic field ensues that could accelerate ions to high energies. This is one of the versions of collective acceleration of ions via high energy electrons that were studied in Professor Rostoker's laboratory in early 1970's in a pursuit of Veksler's vision of collective acceleration of ions. A theory was coined to study this situation [7,8]. Decades later, a nearly similar situation was reincarnated by the advent of intense short pulsed laser irradiation of a foil, from which an energetic electron cloud was ejected out of the foil. Another decade later, as we discussed [21], the irradiation of even thinner foil by even sharper shaped and shorter pulsed laser has now been tried, which enacted the physical situations not so unlike that envisioned in those earlier[7]. Thus we find ourselves in a physical situation that we may be able to embark on some analysis by redeploying the theory that was to explain the earlier physical situation. Theory is thus developed here for ion acceleration in the UHC laser pulse interaction with an ultra-thin (nm-scale) target in which we may be able to regard the coupled electron-ion-electrostatic-field in a set of nonlinear self-similar development. Theory is valid strictly when such is realized. A deceptively simple relationship between the accelerated ion energy and that of electrons is found. We find that the maximum ion energy is $\varepsilon_{max,i} = (2\alpha + 1)Q\varepsilon_{max}$. (This is equal to $(2\alpha + 1)Q[E_0 + \Phi_{pt}]$. when the target is very thin, $\xi \lesssim 1$). This is an increasing function of the coherence parameter $\alpha$, and to the electron energy (The sum of kinetic energy and the ponderomotive potential). The coefficient $\alpha$ is determined by the exponent of backward current density (-J) versus total electron energy. If experimental situations justify the theoretical prerogatives, we find ourselves in an applicability of such a theory to interpret experimental results. In a recent experiment [21,31], we might have had luck to encounter such a case. In the optimum ion acceleration condition $\xi \cong 1$, $\alpha$ is about 3 and 9 for the LP and the CP pulse, respectively. Because of this simplicity, such a formula may be useful in guiding us to anticipate a rough outcome of experiments. That is one of the charms of this theory. When the foil is ultrathin, we recognize that the electron dynamics remains coherent to the laser field. Thus the energies of electrons and ions are assessed directly to the intensity of laser. In this case the above kinetic energy of electrons and the ponderomotive potential energy may be expressed directly by the laser intensity, leading to a further simplified theoretic expression. The theory seems to be predictive for the expansion in this regime. With our model and simulation, it shows that a few Joule laser system with UHC and ultrashort pulse duration (<50 fs) may provide a beam of much more than 250 MeV protons.

When the laser pulse is longer and/or the target is thicker, the physics involves more elements and the parameters that are crucial to the conditions are much highly multidimensional. To sift through this complex physics, we delineate the physical processes into three time stages, the thin layer interaction, the relativistic transparency, and the breakout regime. In another word, for a long pulse with the duration larger than the characteristic time $t_1$, the plasma becomes transparent before the termination of the pulse. It is not without a merit to consider an intellectual exercise, in which the highly nonlinear coupled system of electrons, ions, and electrostatic fields driven by intense laser evolves in a self-similar fashion; under such an event an estimate of ion energy gain may be posted as a guide of experiments (and simulation). In the end what matters most in this regime is the electron energy gain in the breakout afterburner epoch of the laser-electron interaction. Most of the electron energy gain happens during this time. The ion energy gain takes, surprisingly, an analytical form similar to the ultrathin case as a function of electron energy, even though the physical processes to arrive at the electron energy are distinct. The time dependence of the maximum ion energy is also derived, if and when the self-similar evolution is justified. The results are not out of the general behavior of the recent experiments in LANL. It is pertinent to remember that we should remain vigilant about the phenomenological nature of our estimates in these situations and formulaic applications need to be accompanied by caution and wisdom.

When the electron dynamics is slow enough that ions evolve less suddenly, i.e. adiabatically [5,34], the final energy gain of electrons (and thus that of ions) may not be that of the instantaneous energy dictated by the expression $E_0 = m_e c^2 (\sqrt{1 + a_0^2} - 1)$. For example, we have remarked a case of the secondary electron sheet formation that moves together with a class of ions, and a case with a circularly polarized pulse. In the latter, for example, the pulse should cause less electron energy gain than the linearly polarized case so that the cloud of electrons cannot instantaneously shot out of the foil, but more gradually leave the target, rendering a possibility that the electron energy is not only proportional to the field strength ( as proportional to $a_0$), but also to the time over which electrons are accelerated by $\upsilon \times B$ is long enough, to be proportional to (or some fraction of it) $a_0$, leading to the proportionality greater than $a_0$ such as $a_0^2$. This is beyond the scope of the present paper and is left for a future investigation. We anticipate more results to come in advancing the ion energy by laser acceleration spurred by the current theoretical grip of the physics.

### Acknowledgements

This work is supported in part by the DFG Cluster of Excellence MAP (Munich-Centre for Advanced Photonics). XQY would like to thank the support from the Alexander von Humboldt Foundation and NSFC (10855001). TT is supported in part by the Special Coordination Fund (SCF) for Promoting Science and Technology commissioned by the Ministry of Education, Culture, Sports, Science and Technology (MEXT) of Japan. We are deeply indebted by collaboration and discussion with our colleagues F.Krausz, J.Meyer-ter-Vehn, H.Ruhl, A.Henig, D.Jung, D.Kiefer, R.Hoelein, J.Schreiber, J.E.Chen, X.T.He, Z.M.Sheng, Y.T.Li, Z.Y.Guo, J.X.Fang, Y.R.Lu, S.Kawanishi, P. Bolton, S.Bulanov, Y. Fukuda, T. Esirkepov, M.Yamagiwa, M.Tampo, H. Daido, J. Mizuki, M.Abe, M. Murakami,Y. Hishikawa, B.J.Albright, K.J.Bowers, J.C.Fernandez, S.Steinke, M.Schnuerer, T.Sokollik, P.V.Nickles and W.Sandner.